\documentclass[twocolumn,A4,showpacs,preprintnumbers,amsmath,amssymb]{revtex4}
\usepackage{color}
\usepackage{longtable}
\usepackage{graphicx}
\usepackage{dcolumn}
\usepackage{bm}
\definecolor{darkred}{rgb}{0.75,0,0}
\definecolor{blue}{rgb}{0,0,1}

\begin{document}
\bibliographystyle{prsty}

\title{Ionization branching ratio control with a resonance attosecond clock}

\author{Luca Argenti}\email{argenti@physto.se}
\author{Eva Lindroth}\email{lindroth@physto.se} 
\affiliation{Atomic Physics, Fysikum, Stockholm University, AlbaNova University Center, SE-106 91 Stockholm, Sweden}
\date{\today}

\begin{abstract}
We investigate the possibility to monitor the dynamics of autoionizing states in real-time and control the yields of different ionization channels in helium by simulating XUV-pump IR-probe experiments focused on the N=2 threshold. The XUV pulse creates a coherent superposition of doubly excited states which is found to decay by ejecting electrons in bursts. Prominent interference fringes in the photoelectron angular distribution of the $2s$ and $2p$ ionization channels are observed, along with significant out-of-phase quantum beats in the yields of the corresponding parent ions.
\end{abstract}

\maketitle

The evolution of valence electron wave packets in atoms, 
molecules and solids takes place on a timescale ranging 
from tens of attoseconds to few femtoseconds~\cite{Krausz09}. 
For example, the sudden removal of an electron in CO$_2$~\cite{Smirnova09} 
and N$_2$ molecules~\cite{Haessler10} initiates multielectron 
dynamics that unfolds on the attosecond timescale, a localized 
vacancy propagates across the full length of a molecule as large 
as a tetrapeptide~\cite{Remacle06} within just $\sim1$~fs, and
a photoelectron escapes through the surface of solid 
tungsten~\cite{Cavalieri07} in $150$~as or less.
Recent advances in the generation of ultra-short 
pulses~\cite{Sansone06,Goulielmakis08,Sansone08}
provide the tools necessary for a time-resolved 
pump-probe investigation of such dynamics and bear 
the promise of its control.

Most of the reactive processes promoted by high 
electronic excitation, like resonant multiphoton atomic 
ionization~\cite{Nagasono07}, ultrafast 
electron-transfer~\cite{Folisch05}, and molecular dissociative
photoionization~\cite{Martin07,Jiang10,Sansone10},
involve the formation of metastable, multiply excited 
states as a crucial intermediate step.
These metastable states differ from bound excited states in
that they can decay on a timescale that is comparable
to the characteristic time evolution of the electronic
wave packet itself. Their dynamics is an essential
ingredient of the rearrangement of correlated 
multielectron wavefunctions~\cite{Drescher02,Hu05,Uiberacker07}, 
and is thus of particular relevance for their eventual control.
In the present letter, we simulate a realistic 
XUV-pump IR-probe experiment focused on the N=2 ionization 
threshold of helium, the prototype of 
a multielectron system and the simplest neutral atom in which 
autoionizing states arise.
We show that the coherent superposition of doubly excited 
states (DES) created by the XUV pulse ejects electrons in 
bursts rather than continuously, and we demonstrate that it 
is possible to exploit this dynamics to effectively control 
the branching ratios of the different ionization channels.
\begin{figure}[hbtp!]
\includegraphics[scale=0.25]{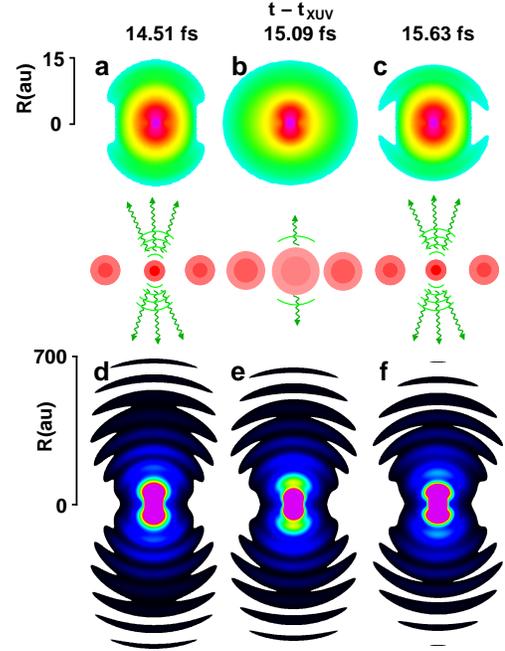}
\caption{\label{fig:breathing} Charge density after the XUV-pump 
pulse, at small (top row) and large (bottom row) radii. 
At each breathing cycle, the metastable wave packet, formed by a 
coherent superposition of doubly excited states, ejects a burst 
of electrons. The peak of the free electron density originating 
close to the nucleus results in a wavefront which propagates 
outward at almost constant speed, up to very large distances.}
\end{figure}


In our simulations, the time-dependent external field comprises 
an XUV-pump pulse followed by an intense IR-probe pulse,
both with a Gaussian envelope. 
The XUV-pump pulse is 385~as long (full width at half maximum
of the intensity), with the energy peaked at $60.69$~eV, and an 
intensity of $2\cdot10^{13}$W$/$cm$^2$. The probe is a 
Ti:Sapphire 800~nm (1.55~eV) pulse, 3.77~fs long (fwhm), with 
an intensity of $10^{12}$W$/$cm$^2$.
The XUV pulse populates a coherent superposition, $|\psi_P\rangle$, 
of {$^1$P$^o$} DES below the N=2 threshold, 
mainly those belonging 
to the principal $sp^+_n$ series~\cite{Fano61}:
\begin{equation}
|\psi_P\rangle\sim\sum_n|sp_n^+\rangle\,c_n\, e^{-iZ_nt/\hbar}
\end{equation}
where $Z_n=E_n-i\Gamma_n/2$ is the complex energy of the 
$sp^+_n$ resonance, with position $E_n$ and width $\Gamma_n$.
The localized part of each term in this series is approximately represented
by a symmetric linear combination of $sp$ configurations,
$sp^+_n\propto 2snp+2pns$~\cite{Fano61}. 
As a consequence, the localized part of $|\Psi_P\rangle$ is 
characterized by a symmetric breathing of $p$ and $s$ orbitals 
coupled to the $2s$ and $2p$ parent ions, respectively:
With the present pulse parameters, the two lowest 
DES in the $sp^+$ series, $sp^+_2$
and $sp^+_3$,
which lie $\sim$5.04~eV and $\sim$1.69~eV below the $N=2$ 
threshold, with lifetimes of $\sim17.6$~fs and $\sim80.3$~fs, 
respectively, are by far the most populated ones.
For several tens of femtoseconds, these two states 
dominate the dynamics of the metastable wave packet.

For the present simulation, the time-dependent
Schr{\"o}dinger equation is integrated numerically with 
an exponential propagator
\begin{equation}\label{eq:prop}
\psi(t+dt)=\exp\left[-i H(t+dt/2)dt/\hbar\right]\psi(t),
\end{equation}
where $H(t)$ is the atomic Hamiltonian in velocity gauge.
The wavefunction $\psi$ is expanded in a multi-channel 
close-coupling B-spline basis with total angular momentum up 
to $L=6$ and the right hand side of Eq.~(\ref{eq:prop}) is 
evaluated with the Arnoldi algorithm. 
Each subspace with definite angular 
momentum $L$ comprises the $1s\phi_L$, $2s\phi_L$,
$2p\phi_{L+1}$, and $2p\phi_{L-1}$ (for $L>0$) 
close-coupling channels, where the notation $nl\phi_{l'}$ 
indicates that one electron is frozen in the $nl$ He$^+$ 
orbital, while the other electron has the orbital 
angular momentum $l'$. 
In the $S$ symmetry, the basis also includes the Hartree-Fock 
$1s_{HF}^2$ configuration for a better representation of the 
ground state.
The radial part of the atomic orbitals is expanded in a 
B-spline basis of order $10$, with an asymptotic spacing between
consecutive nodes of $0.5$ Bohr radii, up to a given maximum 
radius $R$. To compute the yield of the excited ions, a box 
with $R\sim 400$ Bohr radii was found to be sufficient, while for 
the partial differential photoelectron angular distributions 
(PDPAD) a larger box, $R\sim 800$ Bohr radii, was used.
In order to prevent reflections at the box boundaries, a 
channel-specific absorbing potential $V$ is included 
in the Hamiltonian:
\begin{equation}\label{eq:abspot}
V=c\sum_\alpha V_\alpha,\quad V_\alpha=P_\alpha\,\,(r-R_0)^2\,\theta(r-R_0)P_\alpha,
\end{equation}
where the sum runs over all channels, $P_\alpha$ is the 
projector onto the close-coupling channel $\alpha$, $\theta(x)$ is
the Heaviside step function, and $c$ 
is a complex coefficient chosen as $c=-(1+5i)10^{-4}$. 
The radius beyond which the potential is active, $R_0$, is 
set to $\sim100$ Bohr radii from the box boundary.
The absorbing potential $V$ allows one to record the 
annihilation rate in each channel and to reconstruct 
the yields of all the parent ions.
The photoelectron distribution in a channel, identified 
by a parent ion $\alpha=1s,\,2s,\,2p$, is obtained by 
projecting the propagating wave function $\Psi(t)$ onto 
the helium scattering states which satisfy incoming boundary 
conditions in all open channels
but $\alpha$~\cite{Newton}:
\begin{equation}\label{eq:pdcs}
P_\alpha(E,\hat{\Omega})=\sum_{m\sigma\sigma'}
\left|\langle\psi^-_{\alpha,m,\sigma;E,\hat{\Omega},\sigma'}|
\Psi(t)\rangle\right|^2.
\end{equation}
In Eq.~(\ref{eq:pdcs}), $E$ and $\hat{\Omega}$ denote the
photoelectron energy and propagation direction, and the sum runs over the 
projection $m$ of the angular momentum of the electron in 
the parent ion, its spin $\sigma$, and the spin of the 
photoelectron $\sigma'$.
The scattering states are computed with the B-spline
K-matrix method, a well-established configuration interaction 
technique for the single ionization continuum~\cite{Argenti0607}.

In Fig.\ref{fig:breathing}d we show the electron density up to 700 Bohr radii 
at t=14.51 fs after the pump pulse. It consists of distinct 
wavefronts, spreading out with virtually constant speed, separated
by time intervals which correspond closely to the beating period 
between the $sp^+_2$ and the $sp^+_3$ resonances. In other words, the 
metastable wave packet decays by ejecting electrons in isolated 
bursts. This peculiar ``cresting'' behavior~\cite{Robicheaux96}
can be understood in terms of interference between the long range 
part of the wave functions describing the decaying $sp_2^+$ and 
$sp_3^+$ states. A more mechanistic interpretation, however, is
possible.
The Auger decay of DES is known to be triggered 
by electronic correlation; one of the electrons transfers part of 
its excitation energy to the other, which in turn is ejected into 
the continuum.  Pisharody and Jones provided a spectacular and
extreme example of this mechanism~\cite{Pisharody04}; they showed 
that the decay of some autoionizing states of helium, where both 
electrons are 
highly excited, takes place through a single violent e-e collision. A 
similar picture applies also when only one of the two electrons 
is highly excited~\cite{Warntjes00}. In this case, the autoionization 
is found to take place at the encounter of the external electron 
satellite with the excited core. 
In the present case, though, neither of the two electrons is highly
excited. In fact, the metastable wave packet has the smallest 
excitation possible, it lacks a clear semi-classical analogue, 
and the two electrons are constantly in close interaction. 
To investigate whether the collisional point of view still retain 
any validity, we traced the position of 15 consecutive 
wavefronts in the time interval from 10~fs to 30~fs after the 
pump pulse, and extrapolated their evolution backwards in time 
to the moments at which they were created in the vicinity of 
the nucleus. 
The panels in the first and last columns in Fig.~\ref{fig:breathing}
correspond to two selected consecutive times at which a wavefront 
originates close to the nucleus, 14.51~fs and 15.63~fs, while 
the central column corresponds to a time halfway between these two.
In the upper row of Fig.~\ref{fig:breathing} we show the electron 
density within $15$ Bohr radii from the nucleus, which demonstrates
its breathing motion. At $t=14.51$~fs (a) the central part of the wave 
packet is at the peak of its contraction. At $t=15.09$~fs (b) 
it reaches its maximal expansion. Finally, at $t=15.63$~fs (c), 
it is contracted again. The instants at which the wavefronts are 
born in the vicinity of the nucleus therefore correspond closely to the 
stages of maximum contraction of the localized part of the metastable 
wave packet.
This evidence supports the idea that the collisional description 
of the autoionization dynamics of the DES of helium 
is indeed applicable down to the least excited ones.
In the present case, though, it is not the encounter between otherwise 
well-separated electrons~\cite{Pisharody04,Warntjes00} that triggers the 
decay, but rather the squeezing of two electrons in constant close 
interaction.

The XUV pulse has another major effect: it causes the 
sudden ejection of electrons in the $1s$ as well as in 
the $2s$ and $2p$ channels. With the present choice of 
laser parameters, the proportion between the direct 
ionization in the 1s channel, the population of DES and 
the direct ionization in the N=2 channels is roughly 
10:1:0.01. 
\begin{figure}
\includegraphics[scale=0.39]{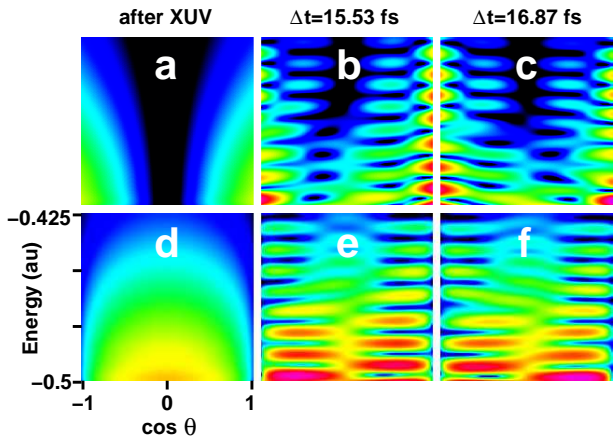}
\caption{\label{fig:2sp}Partial differential 
photoelectron spectra in the $2s$ (top row) and $2p$ 
(bottom row) ionization channels after the XUV-pump 
pulse (a,d) and after the IR pulse with two 
different time delays $\Delta t$ between pump and probe pulses
separated by half the IR period: $15.53$fs (b,e), and 
$16.87$fs (c,f).
x-axis: cosine of the photoelectron ejection angle
with respect to the laser polarization; 
y-axis: photoelectron energy in atomic units.
The interplay between the direct ionization by the XUV and 
the multiphoton ionization of the DES due to the IR-probe
results in prominent interference fringes, with a 
characteristic energy spacing 
$\Delta\epsilon=2\pi\hbar/\Delta t$.}
\end{figure}
In Fig.~\ref{fig:2sp}a and Fig.~\ref{fig:2sp}d we show 
the photoelectron angular distributions in the 
$2s$ and $2p$ channels, respectively, immediately following the
XUV pulse, as functions of both the electron energy (y-axis) 
and the cosine of the angle between the electron 
propagation direction and the polarization of the laser 
(x-axis). In the $2s$ channel, one recognizes the 
characteristic $p$ distribution, proportional to 
$\cos^2\theta$ (the amplitude is odd), while in the $2p$ 
channel the angular distribution results from a combination 
of $s$ and $d$ waves (the amplitude is even).

At the intensity considered, the IR-probe pulse 
has little effect on the ground state. It has a 
profound effect, however, on the DES. 
The population of the $^1$P$^o$ DES is partly 
redistributed among other DES with several different
symmetries, and partly 
promoted to the continuum, mainly to the $N=2$ channels. 
With an intensity of $10^{12}$W$/$cm$^2$, the 
interaction of the system with the IR-probe pulse 
is a typical multiphoton process, where up to four 
IR photons are absorbed. 
As a consequence, the yield of the 2s and the 2p 
parent ions increases roughly by a factor of two,
corresponding to $\sim1\%$ of the population of 
the DES. With more intense probe laser pulses, the 
yield of the excited He$^+$ ions can be substantially 
increased. IR laser pulses with a peak intensity 
of $10^{13}$W$/$cm$^2$ are routinely produced and 
preliminary calculations indicate that, at this 
intensity, the yield of $N=2$ parent ions 
increases by more than one order of magnitude.
\begin{figure}
\includegraphics[scale=0.26]{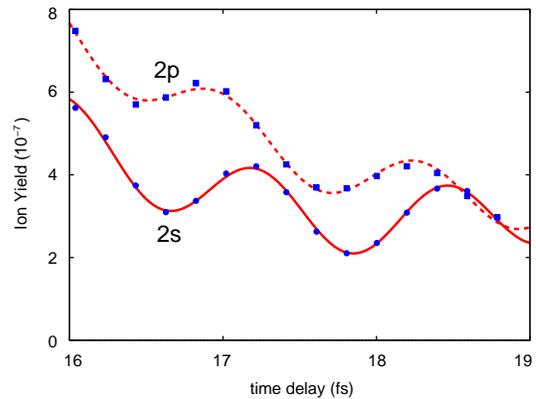}
\caption{\label{fig:osc}Yields of the $2s$ and $2p$ He$^+$
excited parent ions as functions of the time delay 
between the pump and probe pulses. Both yields are modulated 
by large quantum beats, due to the interplay between 
$sp^+_2$ and $sp^+_3$ {$^1$P$^o$} DES, which are out of 
phase by as much as $60^\circ$. The continuous 
curves are obtained by fitting the computed points with 
a sine function plus a quadratic background.}
\end{figure}
The indirect multichannel ionization of DES is interesting
because, by tracking the ionization yields in separate channels, 
one can follow the sharing of both the final energy and angular 
momentum between the two electrons, and hence the 
real-time evolution of electron-electron correlation in 
coherently excited states.
In Fig.~\ref{fig:osc}, the increase in the yields 
of the $2s$ and $2p$ He$^+$ parent ions after the IR pulse 
as functions of the time delay between the two
pulses is reported. Both ion yields are modulated by substantial
quantum beats~\cite{Lange78}. 
A similar phenomenon, due to the coherent superposition 
of bound states rather than resonances, was predicted in 
the ionization of C$^+$~\cite{Lysaght09} and has already 
been observed in helium, close to the N=1 ionization 
threshold~\cite{Mauritsson10}. In the latter case, the 
authors demonstrated that it is possible to control both 
the timing and the probability of ionization.
In the present case, on the other hand, we show that this 
holds also for the branching ratio between 
different ionization channels.
The ion yields track the $sp^+$ breathing mode.
When fitted with the function 
$$
A\cdot\sin(\omega t+\phi)+c_0+c_1 t+c_2 t^2,
$$
the two curves in Fig.~\ref{fig:osc} give the same 
$\omega=0.121(3)$~a.u., which is readily recognized as the energy 
difference between the two lowest $sp^+$ 
resonances, $\Delta E=0.123$~a.u. 
It is interesting to note that the two oscillations 
are out of phase by as much as $60^\circ$, corresponding 
to a time delay of $\sim200$~as; hence, the $2s$ and $2p$ channels 
sample different stages of the breathing motion of the 
metastable wave packet. 
Moreover, the population of the $sp^+_2$ resonance immediately 
after the pump pulse is larger than that of the $sp^+_3$ 
resonance. Since the $sp^+_2$ lifetime is the shortest, there 
is a moment at which the decay rates of the two resonances
become comparable. At this point the beating between the two 
resonances is maximal. These features suggest the possibility 
to control the branching ratio of the two ionization-excitation 
channels and, in turn, to alter the course of reactions 
where metastable electronic states play a dominant role.

There are several ways to detect these quantum beats. 
First, the radiative lifetimes of the 2s and the 2p He$^+$ 
states differ by several orders of magnitude 
($\sim 1.9$~ms~\cite{Prior72} and 
$\sim 10^{-10}$~ms~\cite{Drake92} respectively),
therefore their yields can be disentangled
by looking at their fluorescence decay with ($2s+2p$)
or without ($2p$ alone) the presence of an 
external quenching electric field. 
Second, since the photoelectron angular distribution in the 
$2s$ and $2p$ differ, asynchronous beats should be visible 
in the signal of electrons collected along different directions, 
e.g., along the polarization axis and in the plane orthogonal to it.
The photoelectrons in the N=1 and N=2 channels should give rise 
to two well-separated signals already in a velocity map imaging
spectrometer~\cite{Eppink97}, because of their very different 
energies. If necessary, the latter could be detected in 
coincidence with their excited parent-ion counterpart by 
using a reaction microscope~\cite{Ullrich03}. 

In Fig.~\ref{fig:2sp}e we show the $2p$~PDPAD 
after an IR pulse delayed from the probe by $15.53$~fs.
The probe pulse ionizes the DES generating a short 
series of peaks above the N=2 threshold which interfere 
with the direct-ionization amplitude.
In the lapse between the two pulses, the latter 
accumulates a phase which is linear in both 
the energy and the time delay:
\begin{equation}
\varphi_{direct}(E,t)=\varphi_{direct}(E,t_0)+E(t-t_0)/\hbar.
\end{equation}
As a consequence, prominent interference fringes, 
with characteristic energy spacing 
$\Delta E\sim 2\pi/(t-t_0)$, emerge.
The $sp^+_2$ and $sp^+_3$ $^1$P$^o$ resonances 
are separated from the $N=2$ ionization threshold
by roughly the energy of three and one IR photons, 
respectively.
Since the absorption of an odd number of photons 
by a $^1$P$^o$ state results in an even parity state,
the $2p$ multiphoton ionization amplitude, created
by the IR pulse, should have odd parity just above the threshold 
and change to even parity for photoelectron energies 
around 1.4 eV. Indeed, the interference pattern between 
multiphoton and direct ionization amplitudes (Fig.~\ref{fig:2sp}e)
is asymmetric with respect to $\cos\theta$ close to the threshold, and 
symmetric above $E=-0.45$~a.u. 
At each increase of the time-delay by half an IR cycle, the 
relative phase between direct and indirect multiphoton 
amplitudes changes by $\pi$ close to the threshold, while 
it remains the same one photon energy above.
Indeed, approximately, Fig.2e is the mirror image of Fig.2f.
Similar considerations apply for the $2s$~PDPAD in
Figs.~\ref{fig:2sp}b-c.
By measuring the PDPADs at different time delays~\cite{Mauritsson10}
it is possible to recover the ionization amplitude of the 
DES. With additional information on the phase introduced 
by the IR field in the ionization amplitude of each DES, 
possibly obtained from simulations, even the original 
metastable wave packet could in principle be reconstructed.

In conclusion, we have presented evidence that quantum beating 
between doubly excited states can be monitored experimentally, 
and that it can be exploited with the available attosecond 
pump-probe techniques in order to steer the course of atomic 
photoionization.

We thank Dr.~G.~Sansone, Prof.~A.~L'Huillier and her 
collaborators, and Prof.~H.~Karlsson for useful discussions. 
This work is supported by the G{\"o}ran Gustafsson 
Foundation and the Swedish science research council (VR).

\end{document}